%% file: write-prep.tex
\documentstyle[12pt]{article}

\input feynman

\input{epsf}

\def\ftoday{{\sl  \number\day \space\ifcase\month 
\or Janvier\or F\'evrier\or Mars\or avril\or Mai
\or Juin\or Juillet\or Ao\^ut\or Septembre\or Octobre
\or Novembre \or D\'ecembre\fi
\space  \number\year}}    

\newcommand{\vs}[1]{\rule[- #1 mm]{0mm}{#1 mm}}

\newcommand{\sm}[2]{\frac{\mbox{\footnotesize #1}\vs{-2}}
  {\vs{-2}\mbox{\footnotesize #2}}}
\newcommand{\half}{\frac{1}{2}}
\newcommand{\shalf}{\sm{1}{2}}

\newcommand{\dsla}{\mbox{$\not{\! \partial}$}}
\newcommand{\Dsla}{\mbox{$\not{\! \! D}$}}
\newcommand{\Asla}{\mbox{$\not{\! \! A}$}}
\newcommand{\Zsla}{\mbox{$\not{\! \! Z}$}}

\renewcommand{\a}{\alpha}
\renewcommand{\b}{\beta}
\newcommand{\g}{\gamma}           
\renewcommand{\d}{\delta}         
\newcommand{\e}{\varepsilon}

\newcommand{\la}{\lambda}        
\newcommand{\m}{\mu}

\newcommand{\om}{\omega}         
              
\newcommand{\r}{\rho}
\newcommand{\s}{\sigma}           

\newcommand{\LL}{{\cal L}}
\newcommand{\MM}{{\cal M}}
\newcommand{\OO}{{\cal O}}

\newcommand{\UU}{{\cal U}}

\newcommand{\bfu}{{\mbox {\bf u}}}
\newcommand{\bfd}{{\mbox {\bf d}}}
\newcommand{\bfM}{{\mbox {\bf M}}}
\newcommand{\bfm}{{\mbox {\bf m}}}
\newcommand{\bfS}{{\mbox {\bf S}}}
\newcommand{\bfT}{{\mbox {\bf T}}}
\newcommand{\bfV}{{\mbox {\bf V}}}
\newcommand{\bfC} {{\mbox {\bf C}}}
\newcommand{\bfnu}{{\mbox {\boldmath $\nu$}}}

\newcommand{\sla}{\raise.15ex\hbox{$/$}\kern -.57em} 
\newcommand{\Sla}{\raise.15ex\hbox{$/$}\kern -.70em}

\newcommand{\identity}{\bf 1\hspace{-0.4em}1}
\newcommand{\complex}{{\kern .1em {\raise .47ex
\hbox {$\scriptscriptstyle |$}}
    \kern -.4em {\rm C}}}
\newcommand{\real}{{{\rm I} \kern -.19em {\rm R}}}
\newcommand{\rational}{{\kern .1em {\raise .47ex
\hbox{$\scripscriptstyle |$}}
    \kern -.35em {\rm Q}}}
\renewcommand{\natural}{{\vrule height 1.6ex width
.05em depth 0ex \kern -.35em {\rm N}}}

\newcommand{\twiddle}{\lower.9ex\rlap{$\kern -.1em\scriptstyle\sim$}}

\newcommand{\eq}{\begin{equation}}
\newcommand{\nq}{\end{equation}}
\newcommand{\eqn}[1]{\label{#1}\end{equation}}
\newcommand{\eqa}{\begin{eqnarray}}
\newcommand{\eea}{\end{eqnarray}}
\newcommand{\eqan}[1]{\label{#1}\end{eqnarray}}
\newcommand{\ba}{\begin{array}}
\newcommand{\ea}{\end{array}}
\newcommand{\eqac}{\begin{equation}\begin{array}{rcl}}
\newcommand{\eqacn}[1]{\end{array}\label{#1}\end{equation}}

\newcommand{\remarks}{\bigskip

   \noindent{\bf Remarks:} \begin{enumerate}}
\newcommand{\skramer}{\end{enumerate}}

\begin{document}
\renewcommand{\thefootnote}{\fnsymbol{footnote}}
\newpage
\pagestyle{empty}
\setcounter{page}{0}

\def\logoenslapp{\logolight}
%
%
%
%
\newcommand{\norm}[1]{{\protect\normalsize{#1}}}
\newcommand{\LAP}
{{\small E}\norm{N}{\large S}{\Large L}{\large A}\norm{P}{\small P}}
\newcommand{\sLAP}{{\scriptsize E}{\footnotesize{N}}{\small S}{\norm L}$
${\small A}{\footnotesize{P}}{\scriptsize P}}
\def\logolapin{
  \raisebox{-1.2cm}{\epsfbox{/lapphp8/keklapp/ragoucy/paper/enslapp.ps}}}
\def\logolight{{\bf{{\large E}{\Large N}{\LARGE S}{\huge L}{\LARGE
        A}{\Large P}{\large P}} }}
\begin{minipage}{5.2cm}
  \begin{center}
    {\bf Groupe d'Annecy\\ \ \\
      Laboratoire d'Annecy-le-Vieux de Physique des Particules}
  \end{center}
\end{minipage}
\hfill
\logoenslapp
\hfill
\begin{minipage}{4.2cm}
  \begin{center}
    {\bf Groupe de Lyon\\ \ \\
      Ecole Normale Sup\'erieure de Lyon}
  \end{center}
\end{minipage}
\\[.3cm]
\centerline{\rule{12cm}{.42mm}}

\vspace{20mm}

\begin{center}

{\LARGE {\bf  The Standard Model of particle
physics}\footnote{Lectures presented at the Vth S\'eminaire
Rh\^odanien de Physique, Dolomieu,  March 1997}}\\[1cm]

\vspace{10mm}

{\large P.~Aurenche$^{1}$}\\[.42cm]

{\em Laboratoire de Physique Th\'eorique }\LAP\footnote{URA 14-36 
du CNRS, associ\'ee \`a l'Ecole Normale Sup\'erieure de Lyon et
\`a l'Universit\'e de Savoie.}\\[.242cm]

$^{1}$ Groupe d'Annecy: LAPP, BP 110, F-74941
Annecy-le-Vieux Cedex, France.


\end{center}
\vspace{20mm}

\centerline{ {\bf Abstract}}

\vskip .8cm
\centerline{A brief review of the Standard Model of particle physics
is presented.}
\indent

\vfill
\rightline{hep-ph/9712342}
\rightline{\LAP-A-659/97}
\rightline{July 1997}

\newpage
\pagestyle{plain}
\renewcommand{\thefootnote}{\arabic{footnote}}

\section{Introduction}

From the experimental point of view the world of ``elementary particles"
consists in:\\ 
{\em the leptons}: they have spin $1 \over 2$ and come in three
doublets, $(e^-, \nu_e)$ the electron and its associated neutrino,
$(\mu^-, \nu_\mu)$ the muon and its neutrino, $(\tau^-, \nu_\tau)$ the 
tau and its neutrino.  \\
{\em the vector bosons}: they have spin 1 and there is a massless boson,
the photon, and three massive ones, the $W^+,\ W^-$ and the $Z$. \\
{\em the hadrons}: one distinguishes the mesons of integer spin
($S=0, 1, \dots$) from the baryons of half-integer spin
($S={1\over 2},{3 \over 2},\ \dots$). \\
The {\em leptons} and the {\em vector bosons} are structureless down to
a scale of about $10^{-3}$ to $10^{-4}$ fermi, {\em i.e.} $10^{-18}$ to
$10^{-19}$ m according to the most recent experimental results and they
are treated as elementary fields appearing in lagrangians which describe
the dynamics of their interactions. One the other hand, the {\em
hadrons} have been known for a long time to have a finite size
(typically of the order of 1 fm) and there exist so many hadrons (about
150 mesons and 120 baryons) that they cannot be considered as
elementary. \\
Three types of forces have been identified which describe the dynamics
of particles: the strong force which affects only the hadrons, and the
electromagnetic and weak forces.

The basic principle which guides the construction of models
of particle physics is that of local gauge invariance according
to which the physical properties do not depend on the phases of the
fields. The Standard Model is a (highly successful) example
of a minimal model based on the local gauge group
$$
SU(3) \otimes SU(2)_L \otimes U(1)_Y
$$
{\em i.e.} the direct product of three simple groups.  The dynamical
consequences of this local gauge invariance are  presented in more
details in the next sections. We just state now the main features of
each of these groups.\\
\indent
$-$ The $SU(3)$ gauge group or colour group is the symmetry group of strong
interactions. This group acts on the quarks which are the elementary
constituents of matter and the interaction force is mediated by the
gluons which are the gauge bosons of the group. The quarks and the
gluons are coloured fields. The ``coupling" (fine structure constant)
between quarks and gluons is denoted by $\a_s$ which can be $\ge 1$.
Under some conditions, however, $\a_s$ is very small and perturbation
theory applies. The $SU(3)$ colour symmetry is exact and consequently
the gluons are massless. The theory of strong interactions based on
colour $SU(3)$ is called Quantum Chromodynamics.\\
\indent 
$-$ The $SU(2)_L \otimes U(1)_Y$ is the gauge group of the unified weak
and electromagnetic interactions, where $SU(2)_L$ is the weak isospin
group, acting on left-handed fermions, and $U(1)_Y$ is the hypercharge
group. At ``low" energy ($M < 250$ GeV) the $SU(2)_L \otimes U(1)_Y$
symmetry is ``spontaneously" broken and the residual group is
$U(1)_{emg}$ whose generator is a linear combination of the $U(1)_Y$
generator and a generator of $SU(2)_L$: the corresponding gauge boson is
of course the photon and the associated ``coupling" is {\bf $\alpha$}
$\simeq \frac{1}{137}$. Symmetry breaking implies that the other gauge
bosons acquire a mass: they are the heavy $W^\pm, Z$ bosons dicovered at
CERN in the mid '80's. The symmetry breaking mechanism is associated to
the names of Brout, Englert, Higgs and Kibble, but, for some reasons, it
is often simply referred to as the Higgs mechanism. The  electro-weak
theory, based on spontaneously broken $SU(2)_L \otimes U(1)_Y$ gauge
invariance, is knowned as the Glashow-Salam-Weinberg (GSW) model.

We discuss first the hadronic world and Quantum Chromodynamics (QCD). In
a second part, we study the electro-weak force and the
Glashow-Salam-Weinberg model. 

\section{Quantum Chromodynamics} 

The quark model was introduced more than 30 years ago to describe the
hadrons and we know now that six quarks are necessary, {\em i.e.} the
quarks come in six flavours denoted $q_a = u,d,s,c,b,t$ for quark up,
down, strange, charm, bottom and top.  Like the leptons, the quarks are
point-like, $spin={1\over 2}$ fields so that assuming a baryon is made
up of 3 quarks, $B = (q_a q_b q_c)$, and a meson is made up of a
quark-antiquark pair, $M = (q_a {\overline {q_b}})$, one obtains integer
spins for the mesons and half-integer spin for baryons if account is
taken of the relative orbital angular momentum of the quarks in the
hadrons.  The quarks $d,s,b$ have charge $-{1\over3}$ and the quarks
$u,c,t$ have charge $2\over3$ so that hadrons have only integral
charges. The problem is that a good description of the mass spectrum is
obtained in the quark model with a baryonic wave function which is
symmetrical under interchange of, {\em e.g.}, $q_a \leftrightarrow q_b$:
this violates Fermi  statistics. This problem can be cured if we
introduce a new quantum number, the colour, and assume that each quark
flavour comes in three colours:
\eq
(q_a)^{^T}=(q_a^{1},q_a^{2},q_a^{3})\ \ {\mbox {\rm with}}\ 1,2,3 \ 
{\mbox {\rm the colour indices}}.
\nq
The baryon wave function is then assumed to be totally antisymmetric under
colour interchange,
\eq
B = \sum_{ijk} (q_a^i q_b^j q_c^k),
\nq
and the meson wave function is constructed as
\eq
M = \sum_{i} (q_a^i {\overline {q_b^i}}).
\nq
One introduces a group of transformation in colour space, $(q_a)
\rightarrow (q'_a) = \UU (q_a)$ where $(q_a)$ is written as a column
vector and $\UU$ is a $3\times 3$ matrix. In agreement with all
experimental constraints, the group is chosen to be $SU(3)$, so that
hadronic wave functions are colour singlets ($\identity$), {\em i.e.}
they are invariant under the action of the group. It is postulated that
only colour $\identity$ states are observable which, in agreement with
data, excludes physical states of type $(q_a q_b)$ or $(q_a q_b
\overline q_c)$ (this rules out colour groups such as $O(3)$ or $SO(3)$
for which such unobserved states can also be $\identity$, but does not
rule out $U(3)$ which will be discarded below on dynamical grounds). The
colour hypothesis is further supported by the data on the rate of decay
of $\pi^0 \rightarrow \gamma\gamma$ (via a quark loop) as well as the
rate of hadron production in the high energy $e^+ e^-$ annihilation
reactions (via the process $e^+ e^- \rightarrow q_a^i {\overline
{q_a^i}}$). 

From now on we will take the quarks as the basic degrees of freedom of
the hadronic world. Since quarks are confined into hadrons (no free
quark observed!) one needs a dynamical mechanism to glue the quarks
together. To introduce such a force one proceeds by analogy with Quantum
Electrodynamics (QED) and one makes the colour symmetry  a local
gauge
symmetry. To enforce the invariance of the Lagrangian under a local
$SU(3)$ transformation, it is necessary to introduce eight coloured
vector fields which are in QCD the analogous of the photon in QED. In
more details, the free fermion Lagrangian for one quark field of a given
flavour is
  $$
  {\cal L}=\bar{\psi} (i\not\!\partial -m)\psi
=\bar{\psi}^1 (i\not\!\partial -m)\psi^1 +
\bar{\psi}^2 (i\not\!\partial -m)\psi^2 +
\bar{\psi}^3 (i\not\!\partial -m)\psi^3
  $$
where $\psi$ is a column vector, each coloured component of which is a
Dirac spinor, 
 $
  \psi^{^T}=(\psi^{1{^T}},\psi^{2{^T}} ,\psi^{3{^T}}),
$ $\psi$ belonging to the {\bf 3} (triplet) representation of SU(3). A
local gauge transformation acting on the fundamental {\bf 3} 
representation is parameterised by
  $$
  \UU=e^{ig\sum_{a}\alpha_a(x)T_a},
$$ 
with $g$ the coupling. The $T_a$'s are the eight $3 \times 3$ matrices
generators of the group, and the $\alpha_a(x)$'s are eight arbitrary
real parameters depending on space-time coordinates. The $T_a$'s are
traceless, hermitian and satisfy the commutation relations 
  $
  [T_a, T_b]=if_{abc} T_c,
  $  
with $f_{abc}$ totally antisymetric and real. A gauge
transformation acting on $\psi$ amounts to a locally dependent 
change of phase and for an infinitesimal transformation one has
$$
  \delta \psi= ig\alpha_aT^a\psi, \qquad
  \delta \partial_\mu \psi =ig\alpha_aT^a\partial_\mu \psi
                   +ig(\partial_\mu \alpha_a)T^a\psi, 
 $$  
The Lagrangian is not gauge invariant since its variation is
  \eq
  \delta {\cal L}=\bar{\psi}\left\{ -g(\partial_\mu\alpha_a)
T^a\gamma^\mu \right\} \psi.
  \nq
To regain invariance, it is sufficient to 
introduce 8 vector fields $A_\mu^b(x)$ transforming as
  $$
  \delta A_\mu^c =\partial_\mu\alpha^c -f^{cab}\alpha^a A_\mu^b.
  $$
and consider the lagrangian density
\eq
  {\cal L}=\bar{\psi}\left(i(\partial_\mu-igA_\mu^b(x)T^b)
                  \gamma^\mu -m\right)\psi, 
                  \qquad b=1,\ldots ,8.
\nq
To make the ``gauge bosons" dynamical degrees of freedom
one constructs the anti-symmetric stress-energy tensor
  \eq
 F^a_{\mu\nu}=\partial_\mu A^a_\nu-\partial_\nu A^a_\mu +  g
 f^{abc}A^b_\mu A^c_\nu 
  \label{eq:stress}
  \nq      
and by analogy with QED the ``gauge boson" kinetic term is proportional
to the gauge invariant quantity $F^a_{\mu\nu} F^{a \mu\nu}$. The QCD 
lagrangian density takes then the form 
\eq
\LL_{{\rm QCD}} = - {1 \over 4}  F^a_{\mu\nu} F^{a\mu\nu}\ + \ 
  \bar{\psi}\left(i(\partial_\mu-igA^a_\mu T^a)\gamma^\mu-m\right) \psi
  \label{eq:lqcd}
\nq
where summation over repeated indices is assumed.
Gauge invariance excludes mass terms of the form $m^2_{_A} A_\mu^a A^{a
\mu}$ so that the gauge bosons are massless and induce long distance
forces. We remark at this point that if U(3) had been chosen as the
gauge group one would have a ninth massless gauge boson, which would be
coulourless (associated to the $\identity$ generator). This would induce
long distance strong interactions between hadrons in contradiction with
experiments which tell us that the nuclear force is short-range $\sim
\frac{1}{m_\pi}$. A crucial difference between Chromodynamics and
Electrodynamics is the existence of self-couplings of the gluons. This
is a consequence of the non-abelian nature of the colour group which
imposes the term linear in $f^{abc}$ in eq. (\ref{eq:stress}) and
leads to three-gluon and four-gluon couplings, proportional to $g$ and
$g^2$ respectively, via the kinetic term
$F^a_{\mu\nu} F^{a\mu\nu}$ in eq. (\ref{eq:lqcd}). The consequences of
this fact are crucial and justify the use of the perturbative approach
to study hadronic interactions at high energies as we are now going to
discuss qualitatively. 

When calculating higher order diagrams in field theories like QED or QCD
one encounters ``ultraviolet" divergences, {\em i.e.} divergences at
very high energies or equivalently at very short distances, related to
the point-like structure of fields and the local nature of interactions.
To give a meaning to the perturbative expansion one needs to carry out
the procedure of ``renormalisation" and redefine (renormalize) all the
parameters appearing in the ``bare" Lagrangian eq. (\ref{eq:lqcd}) by
fixing their values at some arbitrary point in momentum space. Expressed
in terms of the renormalized parameters the perturbative series has
finite coefficients. Applied to quark-quark scattering, the matrix
element-squared, {\em i.e.} up to a flux factor the differential cross
section, in the one-loop approximation  (see figure)\\
\vskip .3cm
\centerline{\epsfbox{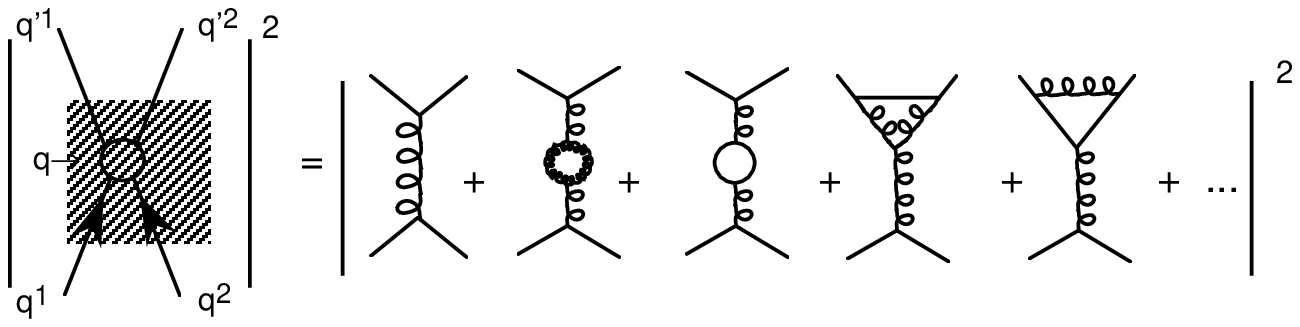}}
\vskip .3cm
\noindent
keeps essentially the same form as at the zeroth order approximation
namely
\eq
| {\MM}^{q_1 q_2 \rightarrow q'_1 q'_2}|^2 \equiv \a_s(q^2)
| \bar{u}(q'_2) \gamma_\mu u(q_2) {1 \over q^2 } 
\bar{u}(q'_1) \gamma^\mu u(q_1) |^2,
\nq 
the main difference with the lowest order result being that the
coupling constant, $\a_s (q^2)= g^2(q^2)/4 \pi$, is running in the sense
that it depends on the momentum scale $q^2$ charateristic of the process
under consideration. More precisely, the calculation leads to 
\eq
\a_s(q^2) = \frac{ \a_s(\mu_0^2)}{1 +  \frac{(11 N_c - 2 N_{_F})}{12
\pi} \  \a_s(\mu_0^2) \ \ln (\frac{-q^2}{\mu_0^2})}  
\label{eq:3.62ter}
\nq 
where $\a_s (\mu_0^2)$ is the renormalised coupling at an arbitrary
energy scale $\mu_0$ and $q^2$ is the momentum transfer of the
scattering process. $N_c = 3$ is the number of colours and $N_{_F}$ is
the number of active flavours . Since for QCD the combination $(11 N_c-
2N_{_F})$ is positive the theory predicts that for large momentum
transfers and therefore large energy the effective QCD coupling
decreases and vanishes at asymptotically high energies: this is called
asymptotic freedom. The smallness of the strong interaction coupling
constant at high energies justifies the use of perturbation theory in
the study of particle physics processes in particular at the colliders such
as those of FERMILAB (proton-antiproton collisions at 1.8 TeV), HERA
(electron-proton collisions at about 300 GeV) or for hadronic proccesses
in $e^+ e^-$ collisions at LEP from 90 GeV to 190 GeV. On the other
hand, at very low energy, the strong interaction coupling is increasing
as the energy decreases and becomes so large as to hopefully lead to the
confinement of quarks and gluons into hadrons at the scale of roughly 1
GeV. Of course, perturbation theory is not applicable then and this
domain is studied using lattice gauge methods. The non-abelian nature of
QCD is crucial to obtain aymptotic freedom as for an abelian theory one
would have $N_c=0$ and the effective coupling would be rising with
energy exactly as it is the case for QED.

The property of asymptotic freedom is not sufficient in itself for
perturbation theory to be useful in the study of hadronic collisions at
high energies. A crucial property of QCD is contained in the
``factorisation theorem" which states that the interaction between hadrons
at  high energies is factorised into a non-perturbative part which
contains the information about the distribution of partons (quarks and
gluons) into the hadrons and a perturbative, calculable part which
describes the interaction between the partons at short  distance. This
is illustrated in the following picture \\
\vskip .3cm
\centerline{\epsfbox{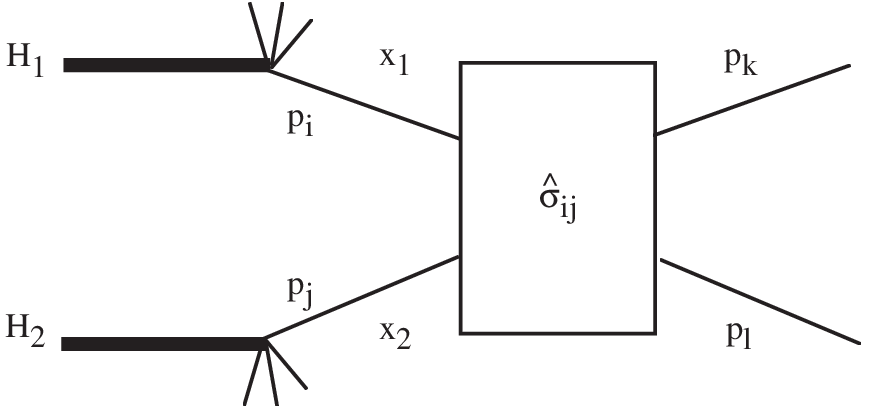}}
\vskip .3cm
\noindent
and made quantitative in the equation giving the generic form of a
scattering cross section between two hadrons
\eqa
d \sigma^{H_1H_2} &=& \sum_{i,j} \int dx_1 \ dx_2 \ F^{H_1}_i (x_1,M) \
F^{H_2}_j(x_2,M) \nonumber \\ && \left[ \hat{\sigma}_{ij} +
\alpha_s(M) \ \hat{\sigma}_{ij}^{(1)} + \alpha^2_s(M)
\hat{\sigma}^{(2)}_{ij} ... \right]. 
\eea 
The structure function (or parton density) $F^{H}_i(x)$ is related to the
probability of finding, in hadron H, a parton of type $i$ carrying a
fraction $x$ of the hadron momentum. The $x$ dependence of $F^{H}_i(x)$
is not calculable in perturbative QCD. The series in square brackets is
the cross section for parton-parton scattering at the energy scale $M$
(or distance $1/M$) charateristic of the process under consideration. It
is the part of the cross section which is calculable provided $M$ is
large enough so that $\a_s(M)$ is small. The theory also predicts the
scale $M$ dependence (scaling violations) of the structure functions.\\
Such cross section have been calculated for most relevant processes up
to the second term in the series and sometimes up to the third term.
A sustained theoretical effort is going on to resum to all orders
large contributions which may appear in the coefficients 
$\hat{\sigma}_{ij}^{(n)}$.

Experimentally there has been for a long time evidence for gluons.
Already in the early deep-inelastic experiments, where one studies  the
scattering of electrons off protons at large momentum transfers, it was
possible to measure the amount of momentum carried by the charged
constituents of the  proton (since the virtual photon exchanged between
the electron and the proton couples only to charged fields). It was
found that charged constituents (the quarks) carry only about half of
the momentum of the proton, the rest being carried by neutral fields (the
gluons?) assumed to bind the quarks together into the hadron. Later on,
with the advent of very high energy colliders it became possible to
``visualise" quarks and gluons as highly  concentrated deposits of
energy, called ``jets" in hadronic calorimeters: these jets fit with the
interpretation of being the hadronic decays of quarks and gluons
produced at large momentum transfers. Perturbative Quantum
Chromodynamics has been tested on many different observables, the most
spectacular of which is probably the jet production cross section as a
function of the jet transverse momentum in proton-antiproton collisions
where experiment and theory agree over ten orders of magnitude. 

For further details one may consult refs. \cite{close}-\cite{pokors}.
For an up-to-date status of QCD phenomenological studies see ref.
\cite{elstiweb}.
\section{The Glashow-Salam-Weinberg model}
Before entering the description of the unified theory of electro-weak
interactions, based on broken gauge invariance, it is useful to briefly
review the Fermi theory of weak interactions and its phenomenological
extensions: this will serve to motivate the choice of the gauge group
$SU(2)_L \otimes U(1)_Y$ as well as illustrate the problems related to
the presence of massive gauge bosons.

\subsection{The Fermi theory and its extensions}
At the beginning was the Fermi theory of neutron decay
$$
n \rightarrow p\ e^-\ \overline \nu_e
$$
(equivalently in the quark model, d quark decay
$d \rightarrow u\ e^-\ \overline \nu_e$)
and muon decay
$$
\mu^- \rightarrow  e^-\ \overline \nu_e\ \nu_\mu.
$$
These transitions were described by a local current-current (4
fermion) interaction parameterised by the Lagrangian
$$
\LL = {G \over {\sqrt 2}} J^\dagger_\a(x) J^\a(x).
$$
The current has a leptonic part and a hadronic part,
$J_\a(x)=l_\a(x)+h_\a(x)$,
\eqa
l_\a(x)&=&\bar{\psi}_e \gamma_\alpha  (1-\gamma_{_5}){\psi_\nu}_e +
  \bar{\psi}_\mu \gamma_\alpha  (1-\gamma_{_5}){\psi_\nu}_\mu + \cdots
 \\
h_\a(x)&=&\bar{\psi}_u \gamma_\alpha  (1-\gamma_{_5})\psi_d +
  \bar{\psi}_c \gamma_\alpha  (1-\gamma_{_5})\psi_s + \cdots
 \label{eq:1}.
\label{eq:curr}
\eea
The particular $V-A$ (vector$-$axial) form of the current, $\g_\a
(1-\g_5)$, is dictated by experiment, in particular the angular
distribution of the decay products. It means that (for massless fields)
only left-handed quarks or leptons $\psi_{_L}=(1-\g_5)\psi/2$ are
sensitive to the weak interactions. The Fermi constant $G$ is universal,
{\em i.e.} it is the same for the hadronic sector and the leptonic sector
and its value has been measured to be
\eq
G = 1.6639 (2) 10^{-5} \ {\mbox {\rm GeV}}^{-2}.
\nq
The transition matrix element for $\mu$ decay is
\eq
\MM = {G \over \sqrt 2} (\bar{\psi}_e \gamma_\alpha 
  (1-\gamma_{_5}){\psi_\nu}_e)(\bar{\psi}_\mu \gamma^\alpha 
  (1-\gamma_{_5}){\psi_\nu}_\mu)^\dagger
\label{eq:fermi}
\nq
which also describes, by crossing symmetry, the scattering process
\eq
\mu^-\ \nu_e  \rightarrow e^- \ \nu_\mu
\nq
at the invariant energy squared $s =(p_{\nu_e} + p_\mu)^2$. The total
cross section for the latter process is then found to be $\s \sim G^2
s$. This result could have easily been guessed on dimensional grounds
since $s$ is the only relevant scale of the problem besides the Fermi
coupling $G$. However such a rapid rise of the cross section with energy
cannot be asymptotically true as it violates the famous Froissart-Martin
unitarity bound which requires $\s \le \ln^2 s$ as $s\rightarrow
\infty$. The problem is related to the locality of the current-current
interaction. To improve the situation one introduces a massive charged
particle to mediate the interaction between the two left-handed
currents. It must be a vector particle because of the $\g_\a$ coupling
in eq. (\ref{eq:curr}).
Denoting $M_{_W}$ the mass of this particle and $g_{_W}$ its dimensionless
coupling to the currents, the matrix element eq. (\ref{eq:fermi}) becomes
\eq
\MM = g^2_{_W} (\bar{\psi}_e \gamma_\alpha (1- \gamma_{_5}) \psi_{\nu_e}) 
{g^{\a\b} - {q^\a q^\b \over M^2_{_W}} \over q^2- M^2_{_W}} 
( \bar{\psi}_\mu \gamma_\beta (1-\gamma_{_5}) \psi_{\nu_\mu} )^\dagger
\label{eq:effcoup}
\nq
where $q^2$ is the momentum transfer of the reaction. At low $q^2$ or
$s$ ($s \ll M^2_{_W}$) one exactly recovers the Fermi model provided one
chooses
\eq
M^2_{_W} =  \sqrt 2 \ {g^2_{_W} \over G}
\nq
while at high energies the total cross section is well behaved since
it is easily derived that
\eq
\s \sim {g^4 \over M^2_{_W}} {s \over s + M^2_{_W}},
\qquad s \gg M^2_{_W}.
\nq
It is useful at this point to comment on the differences between
massless and massive vector bosons concerning the polarisation states.
It is well known that a massless vector particle of momentum
$q_\mu = (q_0,0,0,|q_0|)$, say, has two transverse polarisation (states)
vectors which can be parameterised as 
\eq
\e^\mu_{_T} = (0,1,0,0) \ {\mbox {and}} \ (0,0,1,0) 
\nq
whereas a massive vector particle of momentum
$q_\mu =(q_0,0,0,\sqrt{q^2_0-M^2_{_W}})$ has, besides the two transverse
polarisations, a longitudinal one
\eqa
\e^\mu_{_L} = { 1 \over M_{_W}} (\sqrt {q^2_0-M^2_{_W}},0,0,q_0) 
\sim {q_\mu \over M_{_W}} + \OO ({M_{_W} \over q_0}) 
\label{eq:polar}
\eea
where the rightmost form is asymptotically valid ($|q_0| \gg M_{_W}$).
If $W^\pm$ is a real particle it should be produced in reactions such as
$e^+\ e^- \rightarrow W^+\ W^-$ or, less realistically, in  $\nu_e \
{\overline {\nu_e}} \rightarrow W^+\ W^-$. Considering the latter
\begin{figure}[ht]
\vspace{-.3cm}
\centerline{\epsfbox{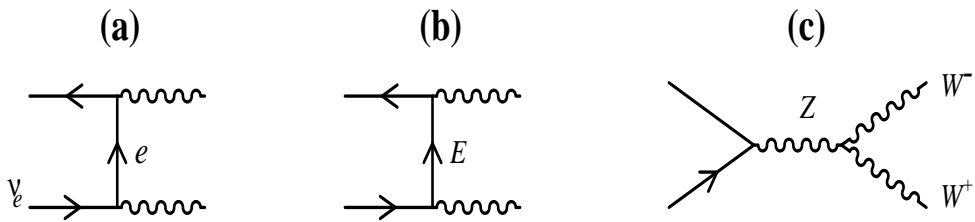}}
\caption{{\em Possible Feynman diagrams for $\nu_e \bar{\nu}_e 
\rightarrow W^+ W^-$ scattering. 
({\bf a}): $e$ exchange;\ ({\bf b}) hypothetical heavy electron $E$
exchange;
\ ({\bf c}) neutral vector boson $Z$ exchange.}} 
\label{fig:2}
\end{figure}
reaction, the amplitude of which is given by only one Feynman diagram
with the exchange of an electron (fig. \ref{fig:2}a), we find the
following results for the production of $W^\pm$ in transverse and
longitudinal polarisation states respectively:
\eqa
  \s (\nu \bar{\nu} \rightarrow W^+_T W^-_T) &\sim& \mbox{constant}
  \nonumber \\
\s (\nu \bar{\nu} \rightarrow W^+_L W^-_L) &\sim&
  {g_{_W}^2 \over M^2_{_W}} \ s
\eea
The origin of the last expression can be understood if one recalls that
the corresponding matrix element is asymptotically
\eq
\MM_{\mu \nu}\ \e^\mu_{_L}(q) \e^\nu_{_L}(q') \sim \MM_{\mu \nu} {q_\mu \over M_{_W}}
{q'_\nu \over M_{_W}}
\nq
leading to the appearence of factors of $s$ in the numerator. We thus
find that the production of longitudinally polarised vector bosons
violates the unitarity limit while that of transverse bosons is well
behaved at high energies. Several ways have been tried to cure this
problem: among them one can mention the hypothesis of a new heavy lepton
(fig. \ref{fig:2}b) and choose its couplings to enforce a proper
behaviour of the cross section at high energies. It turns out that
another possibility, namely that of a heavy neutral vector boson (fig.
\ref{fig:2}c), is realised in Nature. Assuming the most general
couplings ($g_{_Z} \g_\mu (a - b \g_{_5})$, $ g'_{_Z} (p_{_{W^+}} +
p_{_{W^-}})_\mu$), they can be chosen to make cross sections such as $\nu \ e
\rightarrow Z \ W$, $e^+\ e^-\rightarrow W^+ \ W^-, \ \cdots$
asymptotically well behaved. However, this ``bricolage" is not yet
sufficient to have a satisfactory model. Indeed considering $e^+\ e^-
\rightarrow Z \ Z$ scattering and keeping in principle small terms
proportional to the mass of the electron, one finds an interference piece
in the cross section $\sim g^4 m_e \sqrt s / M^4_{_Z}$ which again
violates unitarity! Similar problems arise in the $W\ W$ scattering
process, {\em e.g.} $W^+ \ W^- \rightarrow W^+ \ W^-$ which hopefully
will be studied at LHC or in the future $e^+e^-$ high energy linear
colliders: the cross sections for these processes diverge linearly in $s$.
These  problems can be solved by supposing the existence of a scalar
particle which interacts with the bosons as well as the fermions with
appropriately chosen couplings. 

One can thus construct a viable electro-weak theory in the pedestrian
way described above, carefully choosing masses and couplings of the newly
introduced particles so as to ensure the correct behaviour of all
cross sections. It is more instructive however to assume that these
relations among masses and couplings arise from some hidden symmetry
property. This is what we are going to do next. 
First we describe in some details the symmetry group assuming local
gauge invariance. At this level, the chosen group
requires all fields to be massless. The theory is renormalisable (well
behaved at asymptotic energies) being a non-abelian field theory.
Then, by the mechanism of ``spontaneous symmetry breaking" whereby the
symmetry of the lagrangian is preserved but the choice of a ground state
breaks the symmetry, fermions and gauge bosons acquire a mass.
After symmetry breaking, the theory remains renormalisable as
a consequence of the underlying gauge invariance which imposes the
required relations between couplings. One is left however with a large
number of parameters (at least 19 for the Standard Model) which gives a
motivation for a (still unsuccesful!) search of a deeper symmetry.

\subsection{The $SU(2)_L$ gauge invariance}
As discussed above, the weak interactions induce a transition between
fermions of different charges. It is then natural to group fermions
into doublets
\begin{equation}
    \underbrace{ \left(
      \begin{array}{c}
        \nu_e \\ e^-
      \end{array}
    \right),
    \left(
      \begin{array}{c}
        \nu_\mu \\ \mu^-
      \end{array}
    \right),
    \left(
      \begin{array}{c}
        \nu_\tau \\ \tau^-
      \end{array}
    \right)}_{\mbox{leptons}}\ ;\ 
  \underbrace{ \left(
    \begin{array}{c}
      u \\ d
    \end{array}
  \right),
  \left(
    \begin{array}{c}
      c \\ s
    \end{array}
  \right),
  \left(
    \begin{array}{c}
      t \\ b
    \end{array}
  \right)}_{\mbox{quarks}}
\label{18}
\end{equation}
Because of the $\gamma_\alpha  (1- \gamma_{_5})$  form of the weak
current, we have to distinguish between left-handed fermions which are
sensitive to the weak interactions and the right handed part which are
not. Therefore, each fermion field is decomposed as 
  \eq
    \psi = \frac{1-\gamma_{_5}}{2}\ \psi + \frac{1+\gamma_5}{2}\ \psi 
= \psi_{_L}   + \psi_{_R},  
    \label{19}
  \nq
where $\psi_{_L}$ (resp. $\psi_{_R}$) is identified as the left (resp. 
right) component when $m_\psi=0$. We introduce the left handed doublets:
  \eq
    \psi_{e_L} = \left(
    \begin{array}{c}
      \nu_e \\ e_{_L}
    \end{array}
  \right)
  , \cdots
  \ \ \ \ \ 
  \psi_{q_L} = 
  \left(
    \begin{array}{c}
      u_{_L} \\ d_{_L}
    \end{array}
  \right)
  , \cdots ,
  \label{eq:doublet}
\nq
and the right-handed singlets $\psi_{e_R} = {e_{_R}}, \cdots,
\psi_{q_R} = {q_{_R}}, \cdots$.
The free massless fermion lagrangian is then written 
\eqa
-i  \LL_{F} &=& \bar{\psi}_{e_{_L}} \dsla \ \psi_{e_L} + \cdots +
  \bar{\psi}_{q_L} \ \dsla \ \psi_{q_L} + \cdots +  
\bar{\psi}_{e_{_R}}\ \dsla \ \psi_{e_R} + \cdots +
  \bar{\psi}_{q_R} \dsla \ \psi_{q_R} + \cdots \nonumber \\
&=& \LL_{F_{L}} + \LL_{F_{R}}.
 \label{eq:lfree}
\eea
We impose now a local SU(2) gauge invariance to the left handed part
of the lagrangian, {\em i.e.} we impose $\d \LL_{F_{L}} = 0$ under the
changes of phase
\eq
\delta \psi_{_L} = ig \left( \frac{\vec{\tau}}{2} \cdot \vec{\alpha}(x)
\right)  \psi_{_L} \ \ , \ \ \delta \bar{\psi}_{_L} = -ig \  
\bar{\psi}_{_L}
\left( \frac{\vec{\tau}}{2} \cdot \vec{\alpha }(x) \right)
\nq
where the $2 \times 2$ Pauli matrices $\vec{\tau} = ( \tau_1, \tau_2 ,
\tau_3 )$ satisfy the algebra
\eq
  \left[
    \frac{\tau_i}{2} , \frac{\tau_j}{2} \right] = i \ \epsilon_{ijk} \    
    \frac{\tau_k}{2}. 
\nq
As for the SU(3) local invariance studied above, one has to introduce
a multiplet of gauge vector fields $\vec{W}^\mu(x) = ( W^\mu_1 (x), 
W^\mu_2 (x), W^\mu_3 (x) )$ transforming as
\eq
\delta \ W^\mu_i (x) = \partial^\mu \alpha_i(x) - 
\epsilon_{ijk} \alpha_i(x) \ W^\mu_k (x)
\nq
with $\vec{\alpha }(x) = (\alpha_1(x), \alpha_2(x), \alpha_3(x))$
three real parameters. The locally SU(2) gauge invariant left-handed
lagrangian becomes then
\eq
    \LL_{F_L} =  - \frac{1}{4} \ F^{\mu \nu}_i(x) F_{i_{\mu \nu}}(x)
+ \bar{\psi}_{_L} i \left( \partial^\mu -ig \frac{\vec{\tau}}{2}
    \cdot \vec{W}^\mu(x)\right) \gamma_\mu \psi_{_L} + \mbox{\rm
    {other fermions}}.
\label{eq:lgauge}
\nq
where the antisymmetric tensor $F^{\mu \nu}_i (x)$ is
\eq
    F^{\mu \nu}_i (x) = \partial^\mu W_i^\nu(x) - \partial^\nu 
W_i^\mu(x)\ + g \ \epsilon_{ijk} \ W^\mu_j(x) W^\nu_k(x).
\nq
(see eqs. (\ref{eq:lqcd}), (\ref{eq:stress}). We note at this point the
perfect analogy between the construction of the ``weak"  lagrangian with
that of QCD: the differences are in the choice of group which requires
here only three vector bosons while for $SU(3)$ symmetry eight bosons
had to be introduced. Also, the weak group acts only on the left handed
components of the fields and consequently the $W^\mu_i(x)$ gauge bosons
do not couple to the right handed fermion components. This is called the
weak isospin group and is denoted $SU(2)_L$. The left handed fermion
have isospin $t=1 / 2$ and the third component (eigenvalue of $t_3=
{\tau_3 \over 2}$) is $+1/2$ for $\nu_e, u_{_L}, ...$ and $-1/2$ for
$e_{_L},d_{_L}$. The right-handed components have of course isospin
$t=0$. We assume there are no right-handed neutrinos. Defining the
charged fields $W^{\pm \mu}(x) = (W^\mu_1 \mp i W^\mu_2) / \sqrt{2}$
these two bosons will of course be identified with the charged bosons
introduced in the phenomenological discussion above. One can read off
their interaction term with the fermions from eq. (\ref{eq:lgauge}) and
one finds the familiar ($V-A$) coupling $-i {g \over \sqrt 2} \bar \nu_e
\g^\mu e^-_{_L} W^+_\mu$ and similarly for other fermion doublets (note
the relation ${g / 2 \sqrt 2}= g_{_W}$ of eq. (\ref{eq:effcoup})).
However $W^{3 \mu}$ cannot be the photon nor the $Z$ boson since it also
couples via ($V-A$) in disagreement with data. Anyhow, since we need a
fourth vector boson we introduce a new abelian local invariance and the
point is to show that, by an astute choice of the associated charge, one
can reconstruct the physical couplings. Before doing this let us remark
the three and four boson self-interaction terms buried into the
$F_i\cdot F_i$ term in the lagrangian density eq. (\ref{eq:lgauge})
above.
\subsection{The $U(1)_Y$ gauge invariance}
The $SU(2)_L$ gauge invariance leads to a conserved current: for
example for the  electron sector it is
\eq
J^\mu_3 (x) = \frac{1}{2} \bar{\psi}_{e_L} \gamma_\mu \tau_3 \psi_{e_L}
 = \frac{1}{2} [\bar{\nu} \gamma_\mu \nu - \bar{e}_{_L} \gamma_\mu 
e_{_L}].
\nq 
On the other hand, we know that the electromagnetic current is conserved
\eq
    J^\mu_{emg} (x) =  e_{el} \ \bar{e} \gamma_\mu e = e_{el} \ 
   [\bar{e}_{_L} \gamma_\mu e_{_L} + \bar{e}_{_R} \gamma_\mu e_{_R}],
\nq
where $e_{el}$ is the electron charge expressed in units of the proton
charge ($e_{el}=-1$).
The combination $J^\mu_{emg} (x) - J^\mu_3(x)$ must then necessarily
be conserved. Denoting it the hypercharge current we define
\eq
    \frac{J^\mu_{Y}(x)}{2} \equiv J^\mu_{emg}(x) - J^\mu_3 (x) = -
    \frac{1}{2} \ \bar{\psi}_{e_L} \gamma_\mu \psi_{e_L} - 
    \bar{e}_{_R} \gamma_\mu e_{_R} 
\label{eq:hyper}
\nq
where we recall the definition of the spinor doublet $\psi_{e_L}$ in eq.
(\ref{eq:doublet}). Therefore the hypercharge ${Y_{el} \over 2} = e_{el}
- (t_3)_{el}$ is conserved and its value can be immediately read off
from eq. (\ref{eq:hyper}). One finds the following assignements: $-1$ for
the left-handed doublet and $-2$ for the right-handed electron partner.
The above relation is a particular example of the famous
Gell-Mann/Nishijima formula
\eqa
{Y \over 2} = Q - T_3
\label{eq:gelnish}
\eea which gives the hypercharge in terms of  the electric charge and
the isopin of the field. Associated to this conserved hypercharge we
introduce a local abelian gauge symmetry denoted $U(1)_Y$ to which is
associated a new gauge boson $B_\mu(x)$.  The action of this
transformation on the fermion fields is:
\eqa
      \delta \psi_{_L} = i g' \  y_{\psi_{_L}} \ \beta(x) \ \psi_{_L},
\qquad \delta \psi_{_R} = i g' \ y_{\psi_{_R}} \ \beta(x)\ \psi_{_R}
\eea
with $g'$ the associated coupling constant and the hypercharge values of
the $B_\mu(x)$ field couplings to the fermions are listed in the
following table.
\eq
    \begin{array}{|c|c|c|c|c|}\hline
      & t & t_3 & Y & Q \\ \hline
      \nu_e & 1/2 & 1/2 & -1 & 0 \\ \hline
      e_{_L} & 1/2 & - 1/2 &  -1 & -1 \\ \hline
      e_{_R} & 0 & 0& -2 & -1 \\ \hline
      u_{_L} & 1/2 & 1/2 & 1/3 & 2/3\\ \hline
      d_{_L} & 1/2 & -1/2 & 1/3 & - 1/3\\ \hline
      u_{_R} & 0 & 0 & 4/3 & 2/3 \\ \hline
      d_{_R} & 0 & 0 & -2/3 & -1/3 \\ \hline
\end{array}
\label{eq:table}
\nq
In summary, the initial free lagrangian eq. (\ref{eq:lfree}), becomes
after imposing a $SU(2)$ local symmetry on the left-handed fields
eq. (\ref{eq:doublet}) and an appropriate $U(1)$ invariance on both
the left-handed fields and a right-handed ones,
\eqa
\LL &=& \LL_G \ + \ \LL_F \nonumber \\
&=&  - \frac{1}{4} \ F^{\mu \nu}_i (x) \ F^{\mu\nu}_i (x) -
    \frac{1}{4} \ G^{\mu\nu}(x) \ G^{\mu \nu}(x) \nonumber \\
&& +\ \bar{\psi}_{e_L} \ i \ \Dsla_{_L} \ \psi_{e_L} +
\bar{\psi}_{q_L} \ i \ \Dsla_{_L} \ \psi_{q_L} + \nonumber \\
&& +\ \bar{e}_{_R} \ i \ \Dsla_{_R} \  e_{_R}
+ \bar{u}_{_R} \ i \ \Dsla_{_R}  \ u_{_R} + \bar{d}_{_R} \ i \
\Dsla_{_R}  \ d_{_R}  
\label{eq:lgf}
\eea
where only the electron and $(u,d)$ quark families have been specified.
The left covariant derivative acting on the field $\psi_{_L}$ is 
\eqa
D^\mu_L = \partial^\mu -i \ g \ \frac{\vec{\tau}}{2} \cdot
      \vec{W}^\mu - i \ g' \ y_{\psi_{_L}} \ B^\mu 
\label{eq:dercov}
\eea
and the right covariant derivative acting on field $\psi_{_R}$ 
\eq
    D^\mu_R = \partial^\mu -i\ g' \ y_{\psi_{_R}} \ B^\mu.
\nq
The kinetic term of the gauge fields are constructed from
\eqa
    G^{\mu\nu}(x) &=& \partial^\mu B^\nu(x) - \partial^\nu B^\mu (x)
    \ \ \ \ \ \mbox{(abelian field)} \nonumber \\
    F^{\mu \nu}_i (x) &=& \partial^\mu W^\mu_i(x) - \partial^\nu W^\nu_i(x) + g
    \ \epsilon_{ijk} \ W^\mu_j(x) \ W^\nu_k(x) 
    \label{eq:gaugef}.
\eea 
It is important to point out that $SU(2)_L \otimes U(1)_Y$ invariance
imposes that all fermions are massless. Indeed a fermion mass term in the
lagrangian would have the form
\eq
\LL_{mass} = m\ \bar\psi \psi = m( \bar\psi_{_L} \psi_{_R} \ + 
\ \bar\psi_{_R} \psi_{_L} ).
\nq
But since $\psi_{_L}$ is a doublet and $\bar\psi_{_R}$  a singlet under
$SU(2)$, the mass term cannot be invariant under a gauge transformation!

It is clear from eq. (\ref{eq:hyper}) that the physical fields of 
the photon will be a linear combination of $W^3_\mu$ and the  gauge
boson $B_\mu$ associated to the hypercharge group. Introducing the
fields $A_\mu$ and $Z_\mu$ such that 
\eqa
B^\mu &=& \cos \theta_{_W} \ A^\mu - \sin \theta_{_W} \ Z^\mu \nonumber \\
W_3^\mu &=& \sin \theta_{_W} \ A^\mu + \cos \theta_{_W} \ Z^\mu,
\label{eq:zphot}
\eea
with $\theta_{_W}$ an adjustable parameter, we re-express the relevant
pieces of the lagrangian. Writing out explicitely the result for the 
quark sector we find
\eq
  \LL =
  \left.
    \begin{array}{l}
      \ (\bar{u}_{_L} \bar{d}_{_L}) 
      \left(
        \begin{array}{cc}
          \frac{2}{3} A^\mu & 0 \\
          0 & -\frac{1}{3} A^\mu
        \end{array}
      \right)
      e\ \gamma_\mu 
      \left(
        \begin{array}{c}
          u_{_L} \\ d_{_L}
        \end{array}
      \right)\\
      \\
      + (\bar{u}_{_R} \bar{d}_{_R} ) 
      \left(
        \begin{array}{cc}
          \frac{2}{3}A^\mu & 0 \\
          0 & - \frac{1}{3} A^\mu
        \end{array}
      \right)
      e\ \gamma_\mu 
      \left(
        \begin{array}{c}
          u_{_R} \\ d_{_R}
        \end{array}
      \right)
    \end{array}
  \right\}
  \mbox{photon coupling}
  \label{53}
\nq
\eq
  \mbox{Z coup.}
  \left\{
    \begin{array}{ll}
      + (\bar{u}_{_L} \bar{d}_{_L} ) 
      \left(
        \begin{array}{cc}
          (\shalf - \sm{2}{3} \sin^2 \theta_{_W})Z^\mu
          & 0 \\
&\\
          0 & (- \shalf + \sm{1}{3} \sin^2
          \theta_{_W}) Z^\mu 
        \end{array}
      \right)
      \frac{e}{\sin \theta_{_W} \cos \theta_{_W}} \ \gamma_\mu \left(
      \begin{array}{c}
        u_{_L} \\ d_{_L}
      \end{array}
    \right) \\
&\\
    + (\bar{u}_{_R} \bar{d}_{_R} ) 
    \left(
      \begin{array}{cc}
         (- \sm{2}{3} \sin^2 \theta_{_W})Z^\mu
        & 0 \\
   & \\
     0 & (\sm{1}{3} \sin^2
        \theta_{_W}) Z^\mu 
      \end{array}
    \right)
    \frac{e}{\sin \theta_{_W} \cos \theta_{_W}}\ \gamma_\mu \left(
    \begin{array}{c}
      u_{_R} \\ d_{_R}
    \end{array}
  \right) \\ 
\end{array}
\right.
\label{eq:neutral}
\end{equation}
where the parameters $g, g', \theta_{_W}$ are chosen such that:
\eq
  g \sin \theta_{_W} = g' \cos \theta_{_W}  = e
\label{eq:charge}
\nq
One finds (by construction!) that the $A_\mu$ field has the same couplings
to the two helicity states of a given fermion, therefore it couples
vectorially: we interpret this field as the photon which couples to the
quarks with charges $e_u = {2 \over 3}$ and $e_d = - {1 \over 3}$.
The other neutral field $Z_\mu$, on the other hand, has both
vector and axial-vector couplings to the fermion. The above equation
can be written extremely simply as
\eqa
\LL = e \sum_{q=u,d} e_q \ \bar{q} \ \Asla \ q  \ +\ \frac{e}{\sin
\theta_{_W} \cos \theta_{_W}} \sum_{q=u,d} \bar{q} \ \Zsla \ (a_q + b_q
\gamma_{_5})q  
\label{eq:neutcur}
\eea
with
\eq
a_q = \frac{t_3}{2} -e_q \sin^2 \theta_{_W}, \qquad
b_q = \frac{t_3}{2}.
\label{eq:zcoup}
\nq
A similar derivation can be carried out for the leptonic sector. These
couplings are in agreement with those of the physical $Z$ boson once the
Weinberg angle  $\theta_{_W}$ (in fact introduced by Glashow!) is taken from
experiment to be
\eq 
\sin^2 \theta_{_W} \sim .2322 \ .
\nq
Considering what has been achieved until now, one finds that the model
based on the $SU(2)_L \otimes U(1)_Y$ symmetry contains four gauge
bosons: two charged ones with ($V-A$) couplings to fermions and two
neutral ones with couplings such that these bosons can be interpreted as the
photon and the Z boson. The ``only" difference with the real world 
is that in the present state of development of the model the gauge
bosons are massless, because of the assumed exact gauge invariance
and the fermions are also massless because of the left-right asymmetry
of the gauge group. Counting the bosonic degrees of freedom of the model
one realizes that three degrees of freedom are ``missing", associated to
the longitudinal polarisation states of the heavy vector bosons
as summarized in the table.
\[
  \begin{array}{cc}
    \mbox{\ \ \underline{Model}} &
    \mbox{\ \ \ \ \underline{Real World}} \\
  \mbox{\ \ \ \ degrees of freedom} &  \ \mbox{\ \ \ \ degrees of freedom} \\
    \begin{array}{ccc}
     &\mbox{transverse} & \mbox{longitudinal} \\ 
      W^- & 2 & 0 \\
      W^+ &2&0\\
      Z&2&0 \\
      \gamma &2&0
    \end{array} &
    \begin{array}{ccc}
      &\mbox{transverse} & \mbox{longitudinal} \\ 
      W^- & 2 & 1 \\
      W^+ &2&1\\
      Z&2&1 \\
      \gamma &2& 0
    \end{array}
  \end{array}
\nonumber
\]

In order to complete the model one should therefore introduce at least
three new fields in the lagrangian. This will be done through a
multiplet of scalar fields and it will be seen that, by the mechanism of
spontaneous symmetry breaking of local gauge invariance, some of the
scalar fields become the longitudinal polarization states and
correlatively the vector bosons acquire a mass. The following argument
may help understand the connection between a scalar field and a
longitudinal polarisation state. Consider for example, the $q\bar q Z$
vertex for a boson in a longitudinal state of momentum $q$. It has the
form 
\eq
g\left[ \bar{u} \gamma_\mu (a + b\gamma_{_5})u \right] \epsilon^\mu_{_L}
(q) \sim g \left[ \bar{u} \gamma_\mu (a +\ b \gamma_{_5}) u \right]
  \frac{q^\mu}{M_{_Z}} \ \ \ \ \ |q^\mu|\gg M_{_Z}
\nq
This can be also interpreted, at asymptotic energies, as the derivative 
coupling of a scalar field $\om$ with the fermionic current since, in
momentum space, the derivative brings down a $q^\mu$ factor
\eq
g_\om \left[ \bar{u} \gamma_\mu  (a + b \gamma_{_5})u \right] 
\partial^\mu \om  \leadsto g_\om \left[ \bar{u} \gamma_\mu
(a + b \gamma_{_5})u\right]  q^\mu \om.
\nq
where the coupling $g_\om$ necessarily has the dimension of the inverse
of a mass. In the asymptotic limit a longitudinal boson couples to the
fermions like a scalar with coupling $g_\om \sim g / M_{_Z}$.
\section{Spontaneous symmetry breaking} 
We proceed in steps and discuss, first, the case of a global symmetry
and state the Golstone theorem. It is then  applied specifically to the
electro-weak model. Finally we make the symmetry local which leads us to
the famous Brout-Englert-Higgs-Kibble mechanism. 
\subsection{Global symmetry breaking} 
Consider the very simple case of a complex scalar field
\begin{equation}
\varphi  = \frac{1}{\sqrt{2}} ( \varphi_1 + i \varphi_2) 
\end{equation}
which has two degrees of freedom $\varphi_1(x), \varphi_2(x)$. The
lagrangian is invariant under the global phase transformation
$\varphi(x)  \rightarrow e^{i \alpha} \varphi(x)$ ($\a$ is constant)
where
\eq
\LL = \partial_\mu \varphi^* \partial^\mu \phi - V(\varphi)
\ {\mbox {\rm with the potential }} 
V(\varphi) = -\mu^2 |\varphi|^2 + h |\varphi |^4.
\label{eq:label}
\nq
We assume $\m^2$ is negative so that the potential takes the well-known
``Mexican hat" or ``cul-de-bouteille" shape (depending on your cultural
background!). The hamiltonian is
\eqa
H &=& \pi \ \partial_0 \ \varphi - \LL, \ \ {\mbox {\rm with}}\ \   
\pi = \frac{\delta \LL}{\delta  \partial_0 \varphi} = 
\partial_0 \varphi^* \nonumber \\
&=& \underbrace{\partial_0 \varphi^* \partial_0 \varphi + | \vec{\nabla}
\varphi |^2}_{H_{\mbox{\scriptsize kinetic}}}\ +\ V(\varphi).
\eea
The (positive) kinetic part vanishes for static configurations and the
full hamiltonian is minimal for constant values of the field given by
\eq
|\varphi_0| = \frac{v}{\sqrt2} = \frac{\mu}{\sqrt{2h}} 
\nq
which defines the so-called vacuum expectation value $v$ of the field
$\varphi$ in terms of the parameters of the lagrangian. Indeed, the
quantum theory should be constructed from the lowest energy classical
state which, in this case, is characterised by having its norm
constrained by the above equation: such a state is called the classical
vacuum. One immediately notices that the vacuum is degenerate since
the application of a gauge transformation (phase change) does not affect
the norm of the state. However to construct the quantum theory one needs
to choose a particular vacuum, by imposing, for example, the classical
vacuum field to be real {\em i.e.}
\eq
\varphi_0 = \frac{v}{\sqrt2}
\nq
This obviously amounts to breaking the symmetry of the vacuum since
$\varphi_0$ is no more invariant under a gauge transformation, but the
dynamical laws are still unbroken because they are given by the gauge
invariant lagrangian eq. (\ref{eq:label}). This is the basis of
``spontaneous symmetry breaking" in contradistinction to ``explicit
symmetry breaking" where the lagrangian itself would loose gauge
invariance. To study the theory, we translate the original field
by its vacuum expectation value
\eq
  \varphi = \frac{1}{\sqrt{2}} (v + \varphi_1 (x) + i \varphi_2 (x)) \ \
\nq
and, neglecting constant terms, the lagrangian becomes
\eqa
\LL &=& \frac{1}{2} (\partial_\mu \varphi_1)^2 + \frac{1}{2}
(\partial_\mu \varphi_2)^2 \ -  \ hv^2 \varphi_1^2 \nonumber \\
&&\ + \  h v \varphi_1 (\varphi_1^2 + \varphi^2_2) \ 
+ \ \frac{h}{4} (\varphi^2_1 + \varphi_2^2)^2
\label{eq:lbroken}
\eea
After spontaneous symmetry breaking, we are left with a model
of two interacting real fields $\varphi_1$ and $\varphi_2$. The free
theory is given by the first line of the equation above which shows
that  $\varphi_2$ is massless while $\varphi_1$ has a mass
$m_{\varphi_1} = \sqrt{2hv^2} $. The interaction part is all contained
in the second line of eq. (\ref{eq:lbroken}) and there are cubic and
quartic interactions between $\varphi_1$ and $\varphi_2$. Since, the
initial lagrangian contained only two parameters, there are necessarily
relations between the three parameters $m_{\varphi_1}$, the cubic
coupling $g_{_3}$ and the quartic coupling $g_{_4}$ {\em e.g.}
\eq
g_{_3} = 2 m^2_{\varphi_1} \ g_{_4}.
\nq
These features are a simple illustration of very general properties
of spontaneous breaking of larger (non-abelian) group symmetry.
In particular one can state the Goldstone theorem: \\
\centerline{When a global symmetry is spontaneously broken there appear
as} 
\centerline{many massless scalar modes (called the Goldstone bosons) as 
there}
\centerline{are broken degrees of symmetry.}  

\noindent
A proof of this theorem is now sketched.
Consider a collection of $n$ scalar fields $\varphi_i, \ i=1,\cdots, n$
written as a column vector so that
\eq
\varphi^{^T}= (\varphi_{_1}, \ \cdots, \varphi_{_n})
\nq
The lagrangian density is formally written as
\eq
\LL = \LL(\varphi,\partial_\mu \varphi)_{\scriptstyle {\rm kin}}\ - \ V(\varphi).
\nq
The vacuum of the model is defined by the conditions 
\eq
{\d V \over \d \varphi_i} = 0,\ \leadsto {\mbox{\rm vacuum:}}\ \ 
\varphi^{0{^T}} = (\varphi_{_1}^0, \ \cdots, \varphi_n^0)
\nq
One perturbs around the vacuum state
\eq
\varphi= \varphi^0 + \varphi', \ \ i.e. \ \ 
\varphi_i= \varphi_i^0 + \varphi'_i
\nq
so that the lagrangian (neglecting constant terms) is re-written
\eq
\LL = \LL(\varphi',\partial_\mu \varphi')_{\scriptstyle {\rm kin}}\ 
-\half \sum_{i j} {\d V \over  \d \varphi_i \d \varphi_j}
\varphi'_i \varphi'_j \ \oplus \ (\varphi'^3)\ \oplus \ (\varphi'^4)
\nq
where it is not necessary for our present purposes to specify the cubic
nor the quartic couplings. By construction, there are no terms linear in
the fields because we are expanding around the minimum of the potential.
The quantity of interest is the quadratic term which defines the mass
matrix
\eq 
m^2_{i j} = {\d V \over  \d \varphi_i \d \varphi_j}.
\nq
Consider now the action of an infinitesimal global gauge transformation.
Its action on the fields is
\eq
\d \varphi = i\ \a^J \ T^J\ \varphi,\ \ \ \ J=1, \cdots, N,
\nq where the $T^J$ are the $N$ generators ($n\times n$ matrices) of the
group and the $\a^J$ are the $N$ associated arbitrary parameters. If for
some field configuration $\varphi$ we have for a particular generator
$T^J$,
\eq 
T^J \varphi = 0, \ \ \leadsto \ \d \varphi = i \a^J \ T^J\varphi =0,
\nq
then we say that the configuration $\varphi$ is invariant under the
sub-group generated by $T^J$: the corresponding symmetry is unbroken. If,
on the contrary, $T^J \varphi \ne 0$ the corresponding degree of symmetry is
said to be spontaneously broken. Let us suppose now that the
vacuum satisfies
\eqa
T^J \varphi_0 \ne 0 \ \ && \ \ {\mbox{\rm for}}\ J=1, \cdots, N' \nonumber \\
T^J \varphi_0 = 0 \ \  && \ \ {\mbox{\rm for}}\ J=N'+1, \cdots, N, 
\eea
{\em i.e.} that the vacuum state breaks $N'$ degrees of symmetry. 
The invariance of the lagrangian under the gauge transformation
\eq
\d \varphi_i = i\ \a^J \ T^J_{i k}\ \varphi_k
\nq
for arbitrary $\a^J$
yields ${\d V \over \d \varphi_i} \d \varphi_i \equiv 0$. Taking the
derivative of this relation it comes out
\eq
{\d^2 V \over \d \varphi_j \d \varphi_i} T^J_{i l}\varphi_l \ + \ 
{\d V \over \d \varphi_i} \d_{i j} = 0 \ \leadsto \ 
m^2_{j i}T^J_{i l}\ \varphi^0_l = 0.
\nq
Since this relation is automatically satisfied for $J=N'+1, \cdots, N$
one concludes that the mass matrix must have $N'$ vanishing eigenvalues.
Thus, $N'$ fields $\varphi'_i$ will be massless which are the
Golstone bosons associated to the $N'$ degrees of broken symmetry
({\em qed}).
\subsection{Application to the electro-weak gauge group}
In this case, the generators of the symmetry group will be 
$T^J= (\tau_1,\tau_2,\tau_3, Y)$, {\em i.e.} the
generators of the weak isospin group and of the hypercharge. We
introduce the complex scalar field, which is a doublet of $SU(2)$
($t_{_\Phi} = \half$),
\eqa
\Phi = \frac{1}{\sqrt{2}}
\left(
  \begin{array}{c}
\varphi_{_1} \ -\ i\ \varphi_{_2} \\
 \varphi_{_3} \ -\ i\ \varphi_{_4}
  \end{array}
\right)
\label{eq:phi}
\eea
and the standard scalar lagrangian
\eq
\LL_S = \partial_\mu \Phi^\dag \partial^\mu \Phi -V(\Phi),\ \ 
 V(\Phi) = - \mu^2   \Phi^\dag  \Phi \ +\  h \ (\Phi^\dag \Phi)^2.
\label{eq:lscalar}
\nq
We will define the physical vacuum  by
\eqa
\Phi_0 = 
\left(
  \begin{array}{c}
0 \\
{v \over \sqrt 2}
  \end{array}
\right) \ \ \ {\mbox {\rm with}} \ \ v^2 = {\mu^2 \over h}
\label{eq:vac}
\eea 
such that, of course, $\d V / \d \Phi |_{\Phi=\Phi_0} = 0$. Since we
require the electric charge to be conserved after symmetry breaking, we
have to enforce, following the previous reasoning, that the charge
generator acting on the vacuum state should vanish. Following the
Gell-Mann/Nishijima relation eq. (\ref{eq:gelnish}) we need
\eqa
Q\ \Phi_0 = {1 \over 2}\ (\tau_3 + Y)\ \Phi_0 =
\left(
      \begin{array}{cc}
       \half + {y \over 2} & 0 \\
        0 & - \half + {y \over 2}
       \end{array}
\right) \ \
\left(
      \begin{array}{c}
        0 \\
       {v \over \sqrt 2}
       \end{array}
\right)
\ =\ 0
\label{eq:hyperphi}
\eea 
implying that the hypercharge of the scalar field must be $y_{_\Phi}=1$
to ensure charge conservation in the broken theory. As in the abelian
case, we can study the system around the classical minimum and expand
the scalar field around its vacuum expectation value $v$
\eqa
\Phi = 
\left(
      \begin{array}{c}
       {- i \over \sqrt 2} (\om_{_1}(x) - i \om_{_2}(x)) \\
        {1 \over \sqrt 2} (v + H(x) - i \om_{_3}(x))
       \end{array}
\right) \ = \
\left(
      \begin{array}{c}
        \om^\dag(x) \\
         {1 \over \sqrt 2} (v + H(x) - i \om_{_3}(x))
       \end{array}
\right),
\label{eq:ground}
\eea
so that the scalar potential $V(\Phi)$ becomes
\eqa
V(\Phi) = h v^2 H^2\ +\ h  v \ H  (H^2+\vec\om^2) \ +\ 
\frac{h}{4} \ (H^2+\vec\om^2)^2
\label{eq:pot}
\eea
showing that the triplet of $\om_i$ fields (one neutral and two
charged  fields) is massless while the neutral $H$ field acquires a mass
\eq
M_{_H}= \sqrt{2 h v^2}.
\nq
All these fields are coupled together with a strength which
can be read off the equation above.
\subsection{Local symmetry breaking and the Brout-Englert-Higgs-Kibble
mechanism}
Armed with this lengthy preliminaries we now go directly into the
spontaneous breaking of the local gauge symmetry
$SU(2)_L\otimes U(1)_Y$ down to $U(1)_{\scriptstyle emg}$. Let us state the
results before diving into an ocean of technicalities. The case of a global
symmetry  has just been analysed and led to the appearence of three
massless (Goldstone) bosons  and a massive one. When the symmetry is
made local these massless bosons turn out to be unphysical, in the sense
that they can be gotten rid off by a gauge transformation, but instead,
three gauge bosons (a neutral one and the two charged ones) become
massive and therefore acquire longitudinal polarisation states which
are the Goldstone modes in disguise.

To implement symmetry breaking we first have to extend the electro-weak
lagrangian eq. (\ref{eq:lgf}) to include the scalar field contribution
$\LL_S$ as well as the interaction of the new scalar field with the
fermions $\LL_Y$ (where $Y$ stands for Yukawa) so that the electro-weak
lagrangian is
\eqa
\LL \ = \ \LL_F \ +\ \LL_G\ +\  \LL_S\ +\  \LL_Y.
\label{eq:lgfsy}
\eea
\subsubsection{The scalar lagrangian $\LL_S$}
We concentrate for the moment on $\LL_S$ which drives the spontaneous
breaking of the local electro-weak symmetry. No other fields such as
fermions or gauge bosons can acquire vacuum expectation values otherwise
the physical vacuum would have some angular momentum or other
non-vanishing quantum numbers. The scalar lagrangian has the form
\eqa
\LL_S = |(\partial^\mu -i \ g \ \frac{\vec{\tau}}{2} \cdot
      \vec{W}^\mu - i \ \half \  g' \ B^\mu) \Phi|^2\ -\ V(\Phi)
 \label{eq:ls}
\eea 
where making the symmetry local amounts to replacing the ordinary
derivative in eq. (\ref{eq:lscalar}) by the appropriate covariant
derivative acting on a {\bf 2} field of a local $SU(2)$ symmetry which
also has non-zero hypercharge (see eq. (\ref{eq:dercov})). To analyse
the effects of symmetry breaking we substitute into this equation the
vacuum expectation value $\Phi_0$ of $\Phi$ (eq. (\ref{eq:vac})) and
also work with the  ``physical" $A_\mu$ and $Z_\mu$ fields of eq.
(\ref{eq:zphot}) rather than with $\vec W_\mu$ and $B_\mu$. We have
\eqa
\LL_S |_{\Phi=\Phi_0}\ &=&\ \vline \ 
-{i e \over \sqrt 2 \sin \theta_{_W}} \left(
	\begin{array}{cc} 0 & W^-_\mu \\
			W^+_\mu & 0 
	\end{array} \right)
	\left(	\begin{array}{c}  0 \\
		             {v \over \sqrt 2}
		\end{array} \right) \nonumber \\
& & \  - i e \left(
\begin{array}{cc} 
A_\mu - {\cos^2 \theta_{_W} -\sin ^2 \theta_{_W} \over 2 \sin \theta_{_W} \cos\theta_{_W}}
Z_\mu & 0 \\
0 & - {1\over 2 \sin \theta_{_W} \cos\theta_{_W}}  Z_\mu 
\end{array}    
\right)
 	\left(\begin{array}{c}  0 \\
		 	{v \over \sqrt 2}
		\end{array} \right)
\ \vline^2     -  V(\Phi_0) \nonumber \\
&=& {e^2  \over 4 \sin^2 \theta_{_W}} v^2 W^+_\mu W^{-\mu}\ +\ 
{e^2  \over 8 \sin^2 \theta_{_W} \cos^2 \theta_{_W}} v^2 Z_\mu Z^\mu\ -\ V(\Phi_0)
\label{eq:hboson}
\eea
This equation tells us immediately that the charged bosons $W^\pm$		
acquire the mass $M^2_{_W} = {e^2 \over 4 \sin^2\theta_{_W}} v^2$, the $Z$ boson
a mass $M^2_{_Z} = {e^2 \over 4 \sin^2 \theta_{_W} \cos^2 \theta_{_W}} v^2$, while the
photon remains massless as no quadratic  term in $A_\mu$ appears
in the lagrangian. Note the important relation
\eq
M_{_W} = M_{_Z} \cos\theta_{_W}
\nq
which is a prediction of the model once the boson couplings to fermions
have been determined from experiment. The vanishing of the photon mass
is a consequence of the surviving exact gauge symmetry $U(1)_{emg}$.
We have the relation $v = \frac{\sin\theta_{_W} M_{_W}}{\sqrt{\pi \a}}$
between the vacuum expectation value of the scalar field and the
physical parameters and, plugging in numerical values, we find $v \sim
250$ GeV, which is the basis for the claim, made in the
introduction, that the non-abelian symmetry is broken at the
scale of $250$ GeV.

Turning now to the study of oscillations around the vacuum $\Phi_0$
we write the scalar $\Phi$ in the form eq. (\ref{eq:ground}).
One immediately notices that the three fields $\om_i(x)$ can 
be gauged away since we can always choose the parameters $\vec \a(x)$
of a local gauge transformation such as
\eqa
\exp( - i {\vec \tau \over 2} {\vec \a} (x) ) \  \Phi\ = 
\ \left(\begin{array}{c}  0 \\
		 	{v + H(x) \over \sqrt 2}
		\end{array} \right),
\label{eq:gauge}
\eea
showing that the fields $\om_i(x)$ can be removed from the lagrangian
altogether and therefore are not physical.  This is consistent and, in
fact, it must be so since the massive vector bosons have now a mass and
therefore an extra degree of (longitudinal) polarisation so that the
total number of degrees of freedom is not changed  by the breaking of
the symmetry.  Of course, explicit gauge invariance will be lost since a
particular gauge (the so-called unitary gauge) has been chosen. 
Parameterising the scalar field as in the right-hand side of eq.
(\ref{eq:gauge})  ({\em i.e.} keeping only the physical degrees of
freedom) one immediately reads off from the lagrangian $\LL_S$ the mass
and couplings of the famed Higgs field $H$. From $V(\Phi)$
(see eq. (\ref{eq:pot}) setting $\vec \om = 0$) one gets the mass
\eq
M_{_H} = \sqrt{2h}\ v
\nq
and the triple and quadruple self-couplings respectively
\eq
h_{_3}\sim h v\sim {M_{_H} \over \sqrt h}\sim e {M_{_H}^2 \over M_{_W}},
\ \ {\mbox{\rm and}}\ \ h_{_4} \sim h \sim e^2 {M_{_H}^2 \over M_{_W}^2}.
\nq 
while from the kinetic part of $\LL_S$ come out the couplings of the
Higgs particle to the gauge bosons. To derive them it is enough to make
the substitution $v \rightarrow v+H$ in eq. (\ref{eq:hboson}) and one
obtains \\
$-$ the triple coupling (Higgs-$W^+$-$W^-$) $=  
\frac{e^2}{2 \sin^2\theta_{_W}}  v = \frac{e}{\sin\theta_{_W}} M_{_W}$ \\
$-$ the quadruple coupling (Higgs-Higgs-$W^+$-$W^-$) 
$= \frac{e^2}{2 \sin^2\theta_{_W}}.$\\
The derivation of similar couplings to the $Z$ boson is left as an
exercise.

As discussed before, the existence of the massive gauge bosons leads to a
bad behaviour of the amplitudes at high energies because of the
longitudinal polarisation states. It can be explicitely verified that,
at the tree level, those divergences are cancelled when keeping all
diagrams including those with the Higgs particle. At higher orders,
loop diagrams may involve the massive gauge boson propagator which is of
the form
\eq 
\Delta^{\mu\nu} = i (g^{\mu\nu} - {q^\mu q^\nu \over M^2})/(q^2 - M^2)
\nq
which does not converge to 0 when  $q^\mu \rightarrow \infty$ because of
the  $q^\mu q^\nu / M^2$ term: this leads to an apparently
non-renormalisable theory. As explained in detail for the abelian case
in Piguet's lectures \cite{piguet}, the way out is not to work in the
unitary gauge which keeps only physical degrees of freedom, but rather
to work in a ``renormalisable" gauge (for example, in the 't Hooft
gauge) where the Golstone modes $\vec \om$ are explicitely kept in the
calculation. It can be proven (equivalence theorem) that these modes are
related to the longitudinal polarisation states of the gauge bosons.
\subsubsection{The Yukawa lagrangian $\LL_Y$ and fermion masses and 
couplings} 
The scalar field $\Phi$ can couple to fermions. The requirement for such
couplings to exist is that the corresponding terms in the Lagrangian be
invariant under a $SU(2)_L \times U(1)_Y$ transformation (before the
spontaneous breaking of this symmetry is implemented of course!). Let us
recall that $\psi_{{e_L}}$ and $\psi{_{q_L}}$ of eq. (\ref{eq:doublet})
and $\Phi$ of eq. (\ref{eq:phi}) are {\bf 2} under $SU(2)$ {\em i.e.}
they transform as 
\eq
\d \psi = i\ \frac{\vec{\tau}}{2}\ {\vec{\a}}\ \psi,\ \cdots \ ,
\d \bar\psi = - i\ \bar\psi\ \frac{\vec \tau}{2}\ {\vec \a},\ \cdots  
\nq
so that $\bar\psi_{{e_L}} \Phi$, $\bar\psi_{{q_L}} \Phi$ are invariant
under a $SU(2)_L$ transformation. One easily shows that the combination $ i
\tau_{_2} \Phi^*$ is also a {\bf 2} and consequently the quantites
$\bar\psi_{{e_L}}( i \tau_{_2} \Phi^*)$ and $\bar\psi_{{q_L}}( i
\tau_{_2} \Phi^*)$ are also invariant. Considering now the
transformation properties under $U(1)_Y$ it turns out that the terms
$\bar\psi_{{e_L}} \Phi\ e_{_R}$, $\bar\psi_{{q_L}} \Phi\ d_{_R} $,
$\bar\psi_{{q_L}} (i \tau_{_2} \Phi^*)\ u_{_R}$  are all invariant: this is
easily seen using the fermion hypercharge assignements given in table
(\ref{eq:table})  and the fact that $y_{_\Phi}=-y_{_\Phi{^*}}=1$ (eq.
(\ref{eq:hyperphi})). We do not include a term with $\nu_{_R}$ as we assume
the right-handed neutrino does not couple to the physical world in
agreement with experiment. The  Yukawa lagrangian then takes the form
\eqa
\LL_Y = c_d\ \bar\psi_{{q_{_L}}} \Phi d_{_R} + c_u\ \bar\psi_{{q_{_L}}} (i
\tau_{_2} \Phi^*) u_{_R} + c_e\ \bar\psi_{{e_L}} \Phi e_{_R} + {\mbox{\rm
h.c. + other families,\ \ }} 
\label{eq:yuka}
\eea
where we have explicitely written out the terms involving the first
family of fermions ($\nu, e; u,d$). Six other parameters should be
similarly introduced for the couplings of the second and third families
so that nine new parameters appear in the model. Implementing
spontaneous symmetry breaking, in the unitary gauge, {\em i.e.}
substituting in $\LL_Y$ the expression of $\Phi$ as given in the
right-hand side of eq. (\ref{eq:gauge}), we derive
\eqa
\LL_Y =  
 c_d\ \frac{v+H}{\sqrt 2} {\bar d} d\ +\ 
 c_u\ \frac{v+H}{\sqrt 2} {\bar u} u\ +\
 c_e\ \frac{v+H}{\sqrt 2} {\bar e} e\ +\ {\mbox{\rm other families.\ \ }} 
\label{eq:yukabrok}
\eea
From this expression we relate the mass of a fermion $f$ to the vacuum
expectation value $v$ {\em via}
\eq
M_f = c_f \frac{v}{\sqrt 2}.
\label{eq:massf}
\nq 
This is not a prediction of the theory since the parameters $c_f$ are
unknown and will be adjusted so as to obtain the ``physical" mass of the
corresponding fermion. Furthermore, no relation is expected between
the masses of partners of a given family since one parameter is
introduced  for each of the fermion type in a family. One may remark
that the only ``prediction" is that the neutrino remains massless as a
consequence of the absence a right-handed neutrino. On the other hand
the Higgs couplings to the fermions are predicted, if the fermion masses
are known,
\eq
g_f = \frac{c_f}{\sqrt 2} = \frac{e}{2 \sin\theta_{_W}}\ \frac{M_f}{M_{_W}},
\nq
where eq. (\ref{eq:massf}) and the relation after eq. (\ref{eq:hboson})
have been used: the Higgs particle couples to a fermion flavour in
proportion to its mass, implying that the top quark could play a major
role in the production and/or decay of the Higgs particle ($M_t \sim
175$ GeV) while the electron contribution can be safely neglected ($M_e
= .511\ 10^{-3}$ GeV). 
\begin{figure}[h]
\centerline{\input{ggfus}}
\vspace{-.3cm}
\caption[]{\em  Higgs production mechanism at hadron-hadron colliders.
The dominant contribution arises from a top quark loop.}
\label{fig:3}
\vspace{-.3cm}
\end{figure}
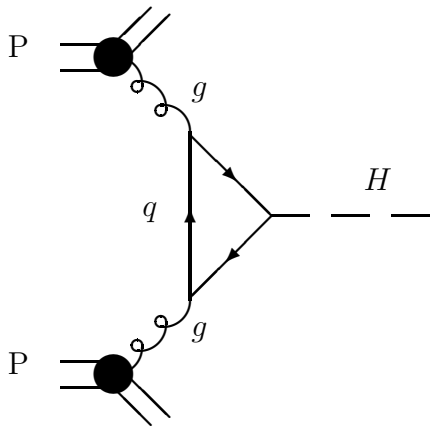
As an application we consider Higgs production in proton-proton
colliders such as LHC (at a center of mass energy $\sqrt s = 14$ TeV).
For a Higgs mass below about 500 GeV the dominant process is gluon-gluon
fusion where the effective Higgs coupling to the gluon-gluon system is
{\em via} a quark loop as indicated in the figure. Because of the very
large top mass the loop contribution is dominated by the top quark. The
possible direct $t \bar t \rightarrow H$ is of course suppressed since
the  density of (virtual) $t, \bar t$ in the proton is negligible. 
\subsection{Family mixing and the Kobayashi-Maskawa matrix}
The above discussion has been considerably simplified since it
completely ignored mixing between generations. However it turns out that
the charged weak interactions are not diagonal in flavour, or to put it
more precisely, the flavour basis defined as the basis which
diagonalises the fermion mass matrix is not the natural one to study
charged current interactions. Furthermore there is no reason for the
absence in the Yukawa lagrangian of terms which mix families. To study
these features let us introduce a new basis of quark flavours {\em e.g.}
($u'_{_L{_j}}$) related to the usual one ($u_{_L{_j}}$) by
\eq
u_{_L{_i}} = S^{u_{_L}}_{i j} u'_{_L{_j}}, \qquad i,j=1,2,3
\nq 
where the indices $i,j$ run over the number of families. This can be
written in a matrix form $\bfu_{_L} = \bfS^{u_{_L}} \bfu'_{_L}$ and
similarly for the right-handed $up$ sector as well as the left-handed
and right-handed $down$ sectors. As will be seen later the matrices
$\bfS^{u_{_L}}, \bfS^{u_{_R}}, \cdots$ are unitary. The quark content
of the $SU(2)_L \otimes U(1)_Y$ lagrangian is written in general
(see eq. (\ref{eq:lgf}))
\eqa
\LL_F = \sum_i (\bar{u}'_{_L{_i}}\ \bar{d}'_{_L{_i}})\ i \ \Dsla_{_L}
\left( \begin{array}{c} 
		{u}'_{_L{_j}} \\
		{d}'_{_L{_j}} 
	\end{array} \right) \ + \
\sum_i \bar{u}'_{_R{_i}}\ i\ \Dsla_{_R} \ u'_{_R{_i}}  \ + \
\sum_i \bar{d}'_{_R{_i}}\ i\ \Dsla_{_R} \ d'_{_R{_i}},\ \ \ 
\label{eq:lmix}
\eea
while the most general Yukawa lagrangian takes the form after symmetry
breaking (see eqs. (\ref{eq:yuka}),(\ref{eq:yukabrok}))
\eqa
\LL_Y &=& \frac{v}{\sqrt 2} \sum_{ij}(
\bar{u}'_{_L{_i}} c^u_{ij} u'_{_R{_j}}\ +\ 
\bar{d}'_{_L{_i}} c^d_{ij} d'_{_R{_j}}\ +\ {\mbox{\rm h.c.}} )
\nonumber \\
&=& \frac{v}{\sqrt 2} ( \bfu'_{_L} \bfC_u \bfu'_{_R}\ +\ 
\bfu'_{_R} \bfC_u^\dag \bfu'_{_L}\ +\ \bfd'_{_L} \bfC_d \bfd'_{_R}\ +\ 
\bfd'_{_R} \bfC_d^\dag \bfd'_{_L} )
\label{eq:yukagen}
\eea
where the complex, $3 \times 3$, $\bfC_u, \bfC_d$ matrices are the
generalised Yukawa couplings. The most general such matrices can be
written as a product
\eq
\bfC_u = \bfM_u\ \bfT_u, \qquad  \bfC_d = \bfM_d\ \bfT_d
\nq
with $\bfM_u$ a hermitian matrix ($\bfM_u=\bfM_u^\dag$) and  $\bfT_u$ a
unitary matrix ($\bfT^{-1}_u=\bfT_u^\dag$). The hermitian matrix can be
diagonalised by a unitary transformation, $\bfM_u = \bfS^{-1}_u \bfm_u
\bfS_u$ where $\bfm_u$ is diagonal with real eigenvalues, and similarly
for the $down$ sector. The Yukawa lagrangian reduces to the very simple
diagonal form
\eqa
\LL_Y &=& \bar{\bfu}_{_L}\ \bfm_u\ \bfu_{_R}\ + \ \bar{\bfu}_{_R}\ \bfm_u\
\bfu_{_L} \ + \ \bar{\bfd}_{_L}\ \bfm_d\ \bfd_{_R}
\ + \ \bar{\bfd}_{_R}\ \bfm_d\ \bfd_{_L}
\nonumber \\
&=& \bar{\bfu}\ \bfm_u\ \bfu\ + \ \bar{\bfd}\ \bfm_d\ \bfd
\eea
when written in terms of the physical basis related to the original one
by
\eqa
\bfu_{_L} &=& \bfS_u \ \bfu'_{_L}, \qquad \bfu_{_R} = \bfS_u\ \bfT_u \ \bfu'_{_R}
\nonumber \\
\bfd_{_L} &=& \bfS_d \ \bfd'_{_L}, \qquad \bfd_{_R} = \bfS_d\ \bfT_d \ \bfd'_{_R}.
\label{eq:rotat}
\eea
We remark that, as advertised before, the transformation from the primed basis to the unprimed
one is unitary since such are the $\bfS, \bfT$ matrices. The components
of $\bfu$  in which the mass matrix is diagonal are, by definition, the
``physical" quark fields $(u,c,t)$ of definite mass eigenstate
({\em idem} for the $d$ sector). Having
achieved a simple form for the Yukawa lagrangian we re-write now the
gauge part $\LL_F$ in terms of these physical fields. Singling out the
neutral current interactions we have
\eqa
\LL_F({\mbox{\rm neutral current}}) = 
\sum_i (\bar{u}'_{_L{_i}}\ i\ \Dsla_{_L} \ u'_{_L{_i}}  \ + \
 \bar{u}'_{_R{_i}}\ i\ \Dsla_{_R} \ u'_{_R{_i}}) \ + \
d'\ {\mbox{\rm sector}},
\eea
in which we keep only the diagonal (in $SU(2)$) part of the operator
$\Dsla_{_L}$. Because of the unitarity of the transformations  within
the left-handed bases and the right-handed bases the above lagrangian
immediately reduces itself to
\eqa
\LL_F({\mbox{\rm neutral current}}) &=&  
\bar{\bfu}_{_L}\ i\ \Dsla_{_L} \ \bfu_{_L}  \ + \
\bar{\bfu}_{_R}\ i\ \Dsla_{_R} \ \bfu_{_R}  \nonumber \\
&+& \bar{\bfd}_{_L}\ i\ \Dsla_{_L} \ \bfd_{_L}  \ + \
\bar{\bfu}_{_R}\ i\ \Dsla_{_R} \ \bfu_{_R}
\eea
This equation is the basis for the slogan that, in the Standard Model
and in agreement with experiments, there is ``no flavour-changing neutral
current", in other words the neutral current is diagonal in flavour space
and does induce transition between the $c$ quark and the $u$ quark or
between the $b$ quark and the $d$ quark for example. The case of the
charged current pieces is more involved because it couples the $up$
sector and the $down$ sector which do not transform with the same
unitary matrices and, as a consequence, there is no reason for the
charged current interactions to be diagonal in the basis which
diagonalises the mass matrix. Indeed we have, from eqs. (\ref{eq:lmix},
\ref{eq:dercov})
\eqa
\LL_F({\mbox{\rm charged current}}) &=&  
g \ \bar{\bfu}'_{_L}\ \g^\mu \ \bfd'_{_L}\ W^-_\mu 
\ + \ {\mbox{\rm h.c.}} \nonumber \\
&=& g \ \bar{\bfu}_{_L}\ \g^\mu \ \bfS_u \bfS_d^\dag\ \bfd_{_L}\ W^-_\mu
\ + \ {\mbox{\rm h.c.}}
\eea
where the last equation is obtained from eq. (\ref{eq:rotat}).
Obviously the matrix $\bfV = \bfS_u \bfS_d^\dag$  is unitary and it
depends therefore on $3^2$ real parameters. Since ($2\times 3 -1$) parameters can be
absorbed in a redefinition of the phases of the six fermion fields
(one overall phase cannot be gotten rid of) there remain four
independent parameters: these can be chosen as three real angles and a
phase and it is convenient to write in the approximate Wolfenstein
parameterisation:
\eq
\bfV = 
\left( \begin{array}{ccc}
1 - \la^2 / 2 		& 	\la  	& A  \la^3(\r - i\eta)	\\
	-\la 		& 1 - \la^2 / 2	&  A  \la^2 \\
A  \la^3(1 - \r - i\eta)& -  A  \la^2 	&	1
					\end{array} \right).
\label{eq:km}
\nq
This is the famous Kobayashi-Maskawa (KM) matrix which generalises to
three families the Cabibbo angle, $\sin\theta_{_{C}} =  \la \sim .22$
introduced long ago to deal with the mixing of two families. This matrix
shows that the charged current transition, for example, of a $d$ quark
to $u,c,t$ quarks takes place with  amplitudes which are proportional to
$(1 - \la^2 / 2), \la , A \la^3(\r - i\eta)$ respectively. The
calculation of processes which are used by experimentalists to measure
accurately the various parameters of the KM matrix and to check its
unitarity  (relations such as $V^*_{ud} V_{ub} + V^*_{cd} V_{cb}+
V^*_{td} V_{tb}=0$ where the $V_{ij}$ are the KM matrix elements) is a
very active area of particle physics phenomenology at present.
The phase factor $\eta$ is responsible for $CP$ violation in the
Standard Model. The measurement of this CP violating parameter, in kaon
and $B$ meson systems, for example, is of great theoretical interest in
order to understand the origin of CP violation and of great practical
importance since it may be related to the origin of the baryon asymmetry
in the universe.

The absence of a KM mixing matrix for the leptonic sector requires a
comment. We assume (and it is supported by experiments) that the
right-handed neutrinos decouple from the observed world. As a
consequence, as mentioned above, the neutrinos $\nu_e, \nu_\mu,
\nu_\tau$ remain massless even after spontaneous symmetry breaking and
there is no mass matrix where the ``physical" neutrino states can be
defined. When studying the weak-current transition from charged leptons
to neutrinos we can introduce a matrix $\bfV_\nu$ similar to the KM
matrix of eq. (\ref{eq:km}) for the quark sector. We are free now
to define the physical neutrino states by
\[
\bfV_\nu^\dag\ \bfnu = \left( \begin{array}{c}
				\nu_e \\
				\nu_\mu \\
				\nu_\tau \end{array}\right)
\]
{\em i.e.} to define the physical states as those for which the 
charged weak current is diagonal in lepton flavour.
\section{Conclusions}
Putting all together, the lagrangian which contains the dynamics of
particle physics, as described by the Standard Model, is decomposed into
\[
\LL= \LL_{QCD} \ + \ \LL_G \ + \ \LL_F \ + \ \LL_S \ + \ \LL_Y 
\]
where each piece has been previously defined. If one attempts to count
the number of parameters introduced we arrive at:

$-$ $SU(3)$ (QCD) gauge invariance: 1 coupling

$-$ $SU(2)\otimes U(1)$ gauge invariance: 2 couplings + the Weinberg
angle $\theta_{_W}$ 

$-$ spontaneous symmetry breaking from $\LL_S$: 2 parameters 

$-$ Yukawa couplings in $\LL_Y$: 9 couplings + 4 KM parameters,

\noindent
so one already has, at least, 19 parameters which is not a satisfactory
situation for a minimal model!  This has prompted an intensive
continuing search for higher hidden symmetries such as supersymmetry,
for example, and (a) Minimal Supersymmetry Standard Model(s) have (has)
been constructed \cite{dereding}, \cite{niles}. The very unfortunate
situation is that in such models the number of fields is more than
doubled compared to the Standard Model as no supersymmetric multiplets
can be filled with only known particles. Furthermore, while in principle
supersymmetry breaking can in turn trigger electroweak symmetry
breaking, one does not know yet how supersymmetry is dynamically
broken. Thus one is lead to describe it in a effective way which
requires many more parameters than in the  Standard Model. It seems
that, at present, the remedy is worse than the disease but there are
still  hopes ......\\

\noindent
For references see {\em e.g.} \cite{quigg,chenli,ryder,peskin}

\vskip .8cm

\noindent
{\large {\bf Acknowledgments}}

I am grateful to F. Boudjema, J.Ph. Guillet and E. Pilon for discussions
and for a careful reading of this text. I also thank F.~Gieres,
M.~Kibler, C.~Lucchesi and O.~Piguet, the organisers of the Vth
S\'eminaire Rh\^odanien de Physique at Dolomieu for a very enjoyable and
educational meeting.

\end{document}

%% file: ggfus.tex


\begin{picture}(20000,17000)
\bigphotons
\THICKLINES
\drawline\fermion[\E\REG](2000,15000)[2000]
\drawline\fermion[\E\REG](2000,14000)[2000]
\put(0,14500){P}
\drawline\fermion[\NE\REG](4000,15000)[2000]
\drawline\fermion[\NE\REG](4000,14000)[3000]
\global\Xone=\pfrontx
\global\Yone=\pfronty
\global\advance \Yone by 500
\global\Ythree=\Yone
\put(\Xone,\Yone){\circle*{1500}}
\drawline\fermion[\E\REG](2000,3000)[2000]
\drawline\fermion[\E\REG](2000,2000)[2000]
\put(0,2500){P}
\drawline\fermion[\SE\REG](4000,3000)[3000]
\drawline\fermion[\SE\REG](4000,2000)[2000]
\global\Xtwo=\pfrontx
\global\Ytwo=\pfronty
\global\advance \Ytwo by 500
\global\Yfour=\Ytwo
\put(\Xtwo,\Ytwo){\circle*{1500}}
\global\negate \Yfour
\global\advance \Ythree by \Yfour
\drawline\gluon[\SE\REG](\Xone,\Yone)[2]
\global\advance \pmidx by 1500
\put(\pmidx,\pmidy){$g$}
\global\multiply \gluonlengthy by 2
\global\advance \Ythree by \gluonlengthy
\drawline\fermion[\S\REG](\gluonbackx,\gluonbacky)[\Ythree]
\drawarrow[\N\ATBASE](\pmidx,\pmidy)
\global\advance \pmidx by -1800
\put(\pmidx,\pmidy){$q$}
\global\multroothalf \Ythree
\drawline\fermion[\SE\REG](\gluonbackx,\gluonbacky)[\Ythree]
\drawarrow[\SE\ATBASE](\pmidx,\pmidy)
\drawline\fermion[\SW\REG](\fermionbackx,\fermionbacky)[\Ythree]
\drawarrow[\SW\ATBASE](\pmidx,\pmidy)
\drawline\scalar[\E\REG](\fermionfrontx,\fermionfronty)[3]
\global\advance \pmidx by 500
\global\advance \pmidy by 1000
\put(\pmidx,\pmidy){$H$}
\drawline\gluon[\SW\REG](\fermionbackx,\fermionbacky)[2]
\global\advance \pmidx by 1500
\put(\pmidx,\pmidy){$g$}

\end{picture}
